\documentclass[aps,pra,twocolumn,floats,showpacs]{revtex4-2}
\usepackage{graphicx}
\usepackage{amsmath}
\usepackage{amssymb}
\usepackage{bm}
\usepackage{color}
\begin{document}

\title{Controlling the degree of rotational directionality in laser-induced molecular dynamics}

\author{Alexander~A.~Milner$^{1}$, and Valery~ Milner$^{1}$}

\affiliation{$^{1}$Department of  Physics \& Astronomy, The University of British Columbia, Vancouver, Canada}

\date{\today}

\begin{abstract}
We demonstrate experimentally a method of varying the degree of directionality in laser-induced molecular rotation. To control the ratio between the number of clockwise and counter-clockwise rotating molecules (with respect to a fixed laboratory axis), we change the polarization ellipticity of the laser field of an optical centrifuge. The experimental data, supported by the numerical simulations, show that the degree of rotational directionality can be varied  in a continuous fashion between unidirectional and bidirectional rotation. The control can be executed with no significant loss in the total number of rotating molecules. The technique could be used for studying the effects of orientation of the molecular angular momentum on molecular collisions and chemical reactions.  It could also be utilized for controlling magnetic and optical properties of gases, as well as for the enantioselective detection of chiral molecules.
\end{abstract}

\maketitle

Intense laser pulses have been long used for controlling molecular rotation (for reviews on this broad topic, see Refs.~\citenum{Stapelfeldt2003,Ohshima2010,Fleischer2012}). The control mechanism is based on the interaction of the electric field of an optical wave with the induced electric dipole of the molecule. The ensuing angle dependent torque, exerted by the field on the molecule, forces the latter to rotate in the direction determined by the parameters of both the field and the molecular polarizability \cite{Zon1975, Seideman1995, Friedrich1995, Vrakking1997, Underwood2005}.

A non-resonant (with respect to the electronic degree of freedom) linearly polarized laser field typically leads to the rotation of the molecular axes toward the polarization vector. As a result, the vectors of the induced molecular angular momentum are distributed uniformly among \textit{all possible directions} in the plane perpendicular to the field polarization. There exists a number of methods to orient the molecular angular momentum along a single axis in the laboratory frame, thus inducing \textit{unidirectional} rotation of molecules. These methods include an application of two consecutive laser pulses \cite{Fleischer2009, Kitano2009}, a chiral pulse train \cite{Zhdanovich2011}, polarization shaped pulses \cite{Karras2015} and an optical centrifuge \cite{Karczmarek1999, Villeneuve2000, Yuan2011, Korobenko2014a}.

In this work, we bridge the gap between the two extreme cases of low and high directionality in the laser-induced molecular rotation. We developed a new method of varying the degree of rotational directionality in a continuous fashion from unidirectional to bidirectional. In the latter case, we control the ratio between the number of molecules rotating clockwise and counter-clockwise with respect to the fixed direction in space. The method is based on modifying the field of an optical centrifuge by adding a variable degree of ellipticity in its field polarization.

\begin{figure}[t]
    \includegraphics[width=0.99\columnwidth]{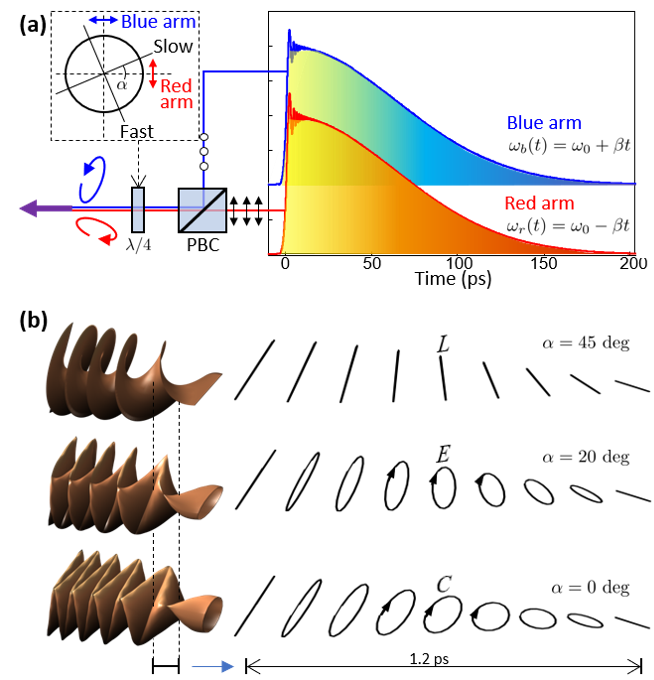}
    \caption{\textbf{(a)} Illustration of the main principle of creating the field of an optical centrifuge. Two oppositely-chirped laser pulses of orthogonal linear polarization are combined on a polarizing beam combiner (PBC) and sent through a quarter-wave plate ($\lambda /4$). Interference of the two beams at the output of the wave plate creates the field of an optical centrifuge. \textbf{(b)} Depending on the orientation of the wave plate with respect to the polarization vectors of the input pulses, expressed via angle $\alpha $ [inset in \textbf{(a)}], the centrifuge polarization undergoes different evolution in time. The three representative examples show, from top to bottom, the behavior of the field polarization over half a period of a ``linear'' (conventional), ``elliptical'' and ``circular'' centrifuge, respectively. The three-dimensional shapes on the left show the surfaces traced by the polarization vector in the course of a few periods of the field evolution. The two horizontal bars at the bottom correspond to 1.2~ps and relate the two types of plots to one another.}
    \label{Fig-cfg}
\end{figure}

A conventional optical centrifuge is a $\approx$100-picosecond long laser pulse, whose polarization remains linear at all times, while the polarization plane is undergoing an accelerated rotation around the propagation direction \cite{Karczmarek1999, Villeneuve2000}. To create such a field, the spectrum of a broadband laser pulse is split in two equal parts (hereafter referred to as ``centrifuge arms'') using a Fourier pulse shaper. The two beams are frequency chirped with an opposite chirp and linearly polarized orthogonally to one another, as illustrated in Fig.~\ref{Fig-cfg}\textbf{(a)}. They are then combined with a polarizing beam combiner and sent through a quarter-wave plate, whose slow optical axis makes an angle of $\pm45$~degrees with respect to the polarization vectors of both input beams. Oriented this way, the wave plate converts the two orthogonal linear polarizations to circular polarizations of opposite handedness, clockwise (CW) for one centrifuge arm and counter-clockwise (CCW) for the other one. Interference of the two circularly polarized laser fields results in a rotating linear polarization, whereas the increasing frequency difference between the two centrifuge arms due to the opposite frequency chirp makes the latter rotation accelerate with time (for specific sets of parameters, see our recent review \cite{MacPhail2020}).

Gas molecules interact with the centrifuge field via the induced electric dipole moment and, if the interaction potential exceeds their thermal energy, follow the accelerated rotation of the centrifuge and reach the state of ultrafast rotation, known as a super-rotor state ($SR$). The process can be described both in classical terms \cite{Karczmarek1999, Armon2016} and as a series of adiabatic quantum transitions through which a molecule climbs the rotational ladder of states \cite{Spanner2001, Spanner2001a, Armon2017}. The efficiency of the rotational excitation, or in other words the capture efficiency of the centrifuge, depends both on the molecular properties (moment of inertia and polarizability anisotropy) and on the field parameters (intensity and angular acceleration). To the best of our knowledge, in all previous theoretical and numerical studies, the field of the centrifuge was assumed to maintain its linear polarization throughout the laser pulse.

On the experimental side, two modifications of an optical centrifuge, affecting its instantaneous polarization, have previously been reported. A so-called ``two-dimensional centrifuge'' was obtained by passing the centrifuge beam through a linear polarizer, therefore constraining the laser polarization to a single 2D plane \cite{Milner2016b}. An oscillating, rather than rotating (as in a conventional three-dimensional centrifuge), linear polarization of such 2D field was shown to excite rotational states as high as the unaltered 3D centrifuge, albeit with lower efficiency. The 2D centrifuge was employed for aligning the molecules in extreme rotational states.

Another approach to modifying the polarization structure of the centrifuge has been taken in \cite{Ogden2019}, where the removal of the quarter-wave plate (equivalent to setting its orientation angle $\alpha $ to zero) produced a field, whose polarization changed from linear to circular and back to linear. It was argued that such a field, dubbed a ``dynamic polarization grating'', does not lead to the high rotational excitations achieved with the rotating linear polarization of a conventional centrifuge.

Here we consider the most general case of a three-dimensional centrifuge, whose polarization structure is determined by a variable orientation of the quarter-wave plate [$-45~\text{deg} \leq \alpha \leq 45~\text{deg}$ in Fig.~\ref{Fig-cfg}\textbf{(a)}]. The conventional centrifuge field is obtained at $\alpha =\pm 45~\text{deg}$, with the direction of its rotation, CW or CCW respectively, determined by the sign of the angle. This is illustrated in the top row of Fig.~\ref{Fig-cfg}\textbf{(b)}. As demonstrated in the middle row, decreasing the absolute value of $\alpha$ below 45~degrees makes the centrifuge polarization elliptical. As before, the polarization ellipse rotates in the direction determined by the sign of $\alpha $, whereas its highest degree of ellipticity (labeled `$E$' in the figure) is defined by $|\alpha |$. When $|\alpha |$ is lowered to zero (equivalent to removing the wave plate, e.g. as in \cite{Ogden2019}), the polarization ellipse ceases to rotate, as depicted in the bottom row of Fig.~\ref{Fig-cfg}\textbf{(b)}. Instead, it oscillates between the two linear states while crossing through the state of circular polarization, labeled `$C$'. As we show in this work, \textit{the centrifuge action of forced accelerated molecular rotation is maintained in all three cases}, i.e. throughout the whole range of angles, including $\alpha=0$. To emphasize this fact, we call the three types of the optical field discussed above as ``linear'', ``elliptical'' and ``circular'' centrifuges, according to the maximum degree of polarization ellipticity attained during its evolution.

Our setup for producing optical centrifuges with different angular acceleration has been described in a recent review \cite{MacPhail2020}. Briefly, we use a regenerative Ti:Sapphire amplifier, which generates laser pulses with 10~mJ energy per pulse and 35~fs pulse length at a central wavelength of 796~nm and a repetition rate of 1~KHz. The centrifuge shaper was constructed following the original design of Villeneuve \textit{et al.} \cite{Villeneuve2000}. The quarter-wave plate at the output of the shaper was mounted on a rotation stage for a continuous scan of the centrifuge ellipticity, as discussed above.

The centrifuge pulses were focused in a cell filled with oxygen gas at room temperature and atmospheric pressure. A focusing lens with the focal length of 10~cm provided the length of the centrifuged region of about 1~mm and a peak intensity of up to $5\times 10^{12}$~W/cm$^{2}$. To characterize the rotation of optically centrifuged molecules, we used polarization-sensitive rotational Raman spectroscopy. After a fixed time delay of 200~ps, the centrifuge was followed by a week circularly polarized probe pulse, derived from the same laser system, spectrally narrowed down to the bandwidth of $0.1$~nm (pulse length of $\sim3$~ps), and frequency doubled to the central frequency of 398~nm. Coherent forward scattering of the probe light by an ensemble of centrifuged molecules results in a rotational Raman shift, whose magnitude is equal to twice the rotational frequency, whereas the sign indicates the direction of molecular rotation with respect to the circular probe polarization \cite{Korech2013, Korobenko2014a}. Throughout this paper, the sense of the circular probe polarization was fixed and coincided with that of the blue centrifuge arm at $\alpha = 45$~degrees.

\begin{figure}[t]
    \includegraphics[width=0.99\columnwidth]{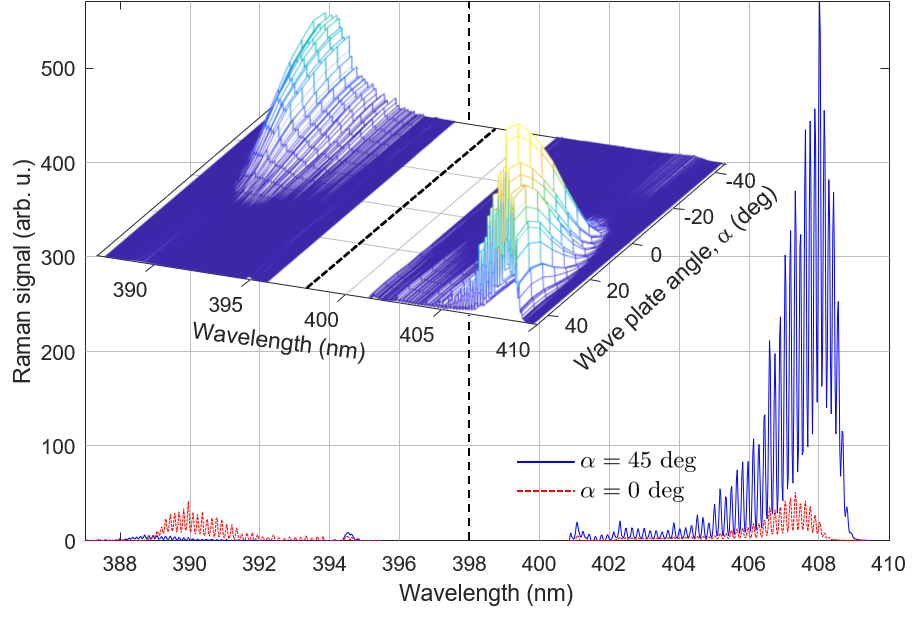}
    \caption{Examples of Raman spectra from the gas of oxygen molecules under ambient conditions, rotationally excited by a conventional linear centrifuge produced by orienting the quarter-wave plate at $\alpha =45$~degrees (solid blue), and a circular centrifuge created by setting $\alpha $ to zero degrees (dashed red). Sharp peaks correspond to separate rotational quantum states of oxygen molecules. The three-dimensional inset shows a full continuous evolution of the Raman spectrum from purely Stokes (positive wavelength shift) at $\alpha =45$~deg to purely anti-Stokes (negative wavelength shift) at  $\alpha =-45$~deg through intermediate values of the centrifuge ellipticity at $\alpha \neq \pm45$~deg, where it contains both Stokes and anti-Stokes components. Dashed lines at 398~nm mark the central probe wavelength.}
    \label{Fig-raman}
\end{figure}

Fig.~\ref{Fig-raman} shows Raman spectra of the centrifuged oxygen with well-resolved peaks, corresponding to the individual rotational quantum states. When the molecules rotate in the same direction as the field vector of the circularly polarized probe pulses, the spectrum consists of the frequency-downshifted Stokes lines, depicted by the solid blue curve. In the opposite case of all molecules counter-rotating with respect to the probe polarization (not shown), only upshifted anti-Stokes peaks show up on the other side of the central probe frequency. The appearance of both Stokes and anti-Stokes lines reflects an intermediate situation, when part of the molecular ensemble rotates in the direction of the probe polarization, and another part - in the opposite one. An example is shown in Fig.~\ref{Fig-raman} with the dashed red line, corresponding to the circular centrifuge produced with $\alpha =0$ (CW if observed along the beam).

The dependence of the measured Raman spectrum on the angle of the quarter-wave plate is plotted as a 3D inset in Fig.~\ref{Fig-raman}. One can see that when $\alpha $ changes from $+45$ to $-45$~degrees, Stokes peaks are falling in amplitude while anti-Stokes are gaining strength. Notably, the two centers of mass of both Stokes and anti-Stokes manifolds remain almost constant, indicating that an optical centrifuge with \textit{any degree of ellipticity} is exciting the molecules to the similar rotational frequencies, as those obtained with the conventional linear centrifuge.

Originating from a coherent scattering effect, peak amplitudes in the recorded Raman spectra are proportional to the square of the number of molecules, rotating with the corresponding (to the Raman shift) frequency \cite{Bitter2016c}, weighed by the respective Raman transition strength. The latter is determined by the two-photon interaction matrix elements, proportional to the $3j$ symbols \cite{Seideman2005}
\begin{equation}\label{eq-3j}
\mathcal{W}_{J,m} = \left( \begin{array}{ccc} J & |\Delta J| & J+\Delta J \\ m & \Delta m & -(m+\Delta m) \end{array} \right),
\end{equation}
\noindent where $J$ ($m$) are the rotation (magnetic) quantum numbers, $\Delta J=2$ $(-2)$ corresponds to a Stokes (anti-Stokes) transition, and $\Delta m=2$ $(-2)$ is associated with the right (left) circular polarization of the scattered photons.

\begin{figure}[t]
    \includegraphics[width=0.99\columnwidth]{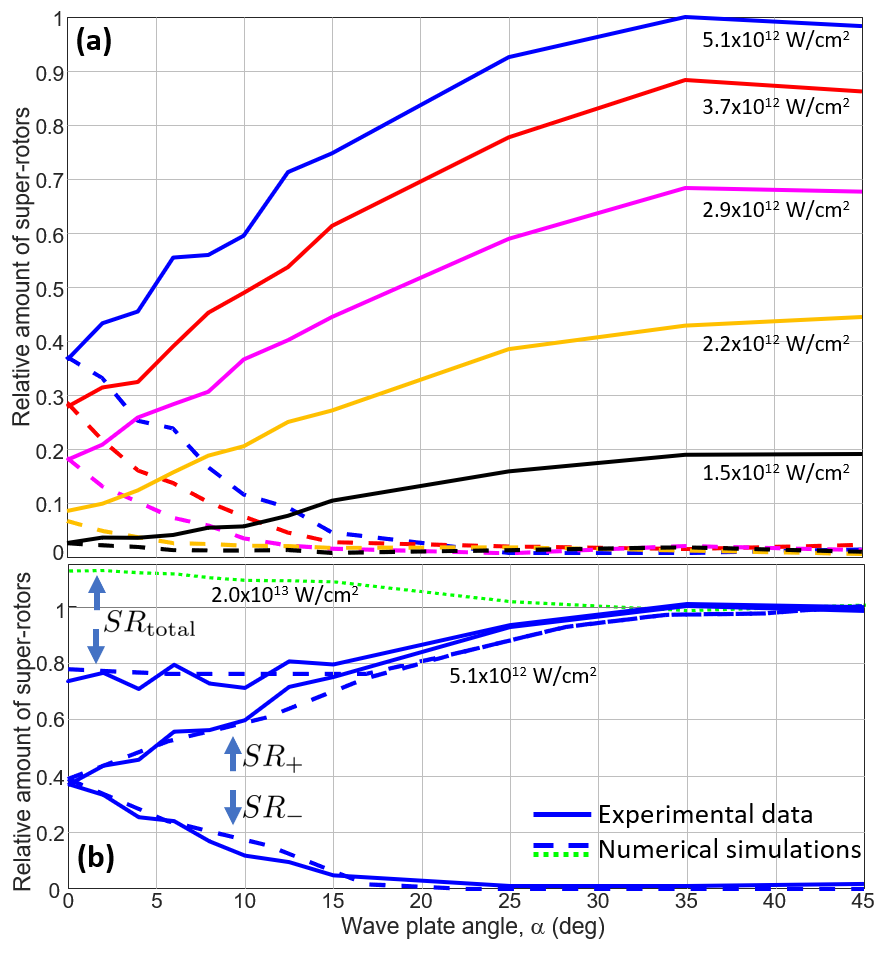}
    \caption{\textbf{(a)} Experimentally determined relative amount of ``positive'' super-rotors $SR_+$ (i.e. rotating in the direction of the centrifuge, solid lines) and ``negative'' super-rotors $SR_-$ (rotating against the centrifuge direction, dashed lines) as a function of the orientation angle of the quarter-wave plate $\alpha $ in the centrifuge shaper. The number of super-rotor was calculated from the Raman spectra recorded at five different peak intensities of centrifuge pulses, indicated under the right hand side of each solid curve. Solid $SR_+$ lines connect to the corresponding dashed $SR_-$ lines at $\alpha =0$, where by the symmetry of a ``circular'' centrifuge, the numbers of CW and CCW rotors are equal to one another ($SR_+ \equiv SR_-$). \textbf{(b)} Comparison of the experimental data for the peak intensity of $5.1\times 10^{12}$~W/cm$^2$ (blue solid lines) with the result of numerical simulations (blue dashed lines). The curves representing the total number of super-rotors are labeled with $SR_\text{total} (\equiv SR_+ + SR_-)$. The dotted green line at the top shows the calculated value of $SR_\text{total}$ for the centrifuge intensity of $2.0\times 10^{13}$~W/cm$^2$.}
    \label{Fig-populations}
\end{figure}

From the properties of $3j$ symbols at $J\gg 1 \text{ and } |m|\approx J$, it follows that if the molecular ensemble consisted of molecules rotating both in the direction of the centrifuge ($m\approx J$, hereafter referred to as ``positive'' super-rotors) and against it ($m\approx -J$, ``negative'' super-rotors), the former would produce predominantly Stokes photons, whereas the latter would contribute mainly to anti-Stokes scattering. This separation of frequencies means that no optical interferences would exist between the Raman-shifted light originating from positive and negative super-rotors, greatly simplifying the conversion of Raman amplitudes to the amount of molecules spinning in each direction. We also note that since the widths of the rotational wave packets, created by the centrifuge, both in $J$ and $m$, are much smaller than the corresponding average values $\overline{J} \text{ and } \overline{m}$, we further simplify the analysis and assume an equal weight of all Raman lines in Eq.~(\ref{eq-3j}). That is, when calculating the approximate population of each rotational state, we take
\begin{equation}\label{eq-approx3j}
\mathcal{W}_{J,m} \equiv \left( \begin{array}{ccc} \overline{J} & |\Delta J| & \overline{J}+\Delta J \\ \overline{m} & \Delta m & -(\overline{m}+\Delta m) \end{array} \right)
\end{equation}
\noindent for all $J$ and $m$ belonging to the same rotational wave packet. Fig.~\ref{Fig-populations}(\textbf{a}) shows the numbers of oxygen super-rotors determined from the experimental Raman spectra and plotted as a function of the quarter-wave plate angle $\alpha $ for different centrifuge intensities. The amount of super-rotors is normalized to the maximum value among all data points. One can see that when $\alpha $ decreases from 45 to about 20~degrees, the number of molecules spun in the direction of the centrifuge ($SR_+$, solid lines) falls down. This behavior is well expected due to the increasing ellipticity of the centrifuge field, and correspondingly lower intensity of its linearly polarized component, accompanied by the decreasing capture efficiency. Interestingly, when $\alpha $ drops below $\approx 20$~degrees, the appearance of anti-Stokes lines in the Raman spectrum (see inset in Fig.~\ref{Fig-raman}) indicates the emergence of negative super-rotors ($SR_-$, dashed lines), whose number grows to a significant level of up to 40\%. Clearly, \textit{an elliptical centrifuge is capable of spinning the molecules in both directions}. In the limiting case of $\alpha =0$, a circular centrifuge produces an equal amount of super-rotors with two opposite senses of rotation.

Note that at high intensities, the maximum number of positive super-rotors is reached at $\alpha =35$~degrees, instead of an anticipated 45~degrees. We attribute a small rise in numbers, when $\alpha $ changes from 45 to 35~degrees, to the nonlinear birefringence (Kerr effect) in the optical components of our setup, which may affect the polarization of the centrifuge field, making it more linear at $\alpha =35$ rather than 45~degrees.

To compare our experimental results with the theoretical expectations, we carried out numerical simulations of the classical interaction of oxygen molecules with the field of an optical centrifuge. The calculation procedure was based on numerically solving Euler equations for the angular frequency $\bm{\Omega}$ of a molecule with the moment of inertia $\hat{\mathbf{I}}$ subject to the torque $\mathbf{T}$ \cite{Tutunnikov2018},
\begin{equation}\label{eq-Euler}
\hat{\mathbf{I}} \dot{\bm{\Omega}} = \left( \hat{\mathbf{I}}\bm{\Omega} \right) \times \bm{\Omega} + \mathbf{T}.
\end{equation}
\noindent For the initial conditions, we took a thermal ensemble of 10,000 molecules and assumed an isotropic distribution of molecular axes and Maxwell-Boltzmann distribution of angular velocities. The centrifuge-induced torque was calculated as described below.

The field of an optical centrifuge with an arbitrary degree of ellipticity, i.e. for an arbitrary angle $\alpha $ of the quarter-wave plate in the centrifuge shaper, can be calculated by starting with the field vector before the wave plate,
\begin{equation}\label{eq-E0}
\mathbf{E}_0 \equiv \left[ \begin{array}{c} E_x \\ E_y \end{array} \right] = \mathcal{E}_0 e^{-i\omega_0 t} \left[ \begin{array}{c} e^{-i\varphi(t)} \\ e^{+i\varphi(t)} \end{array} \right],
\end{equation}
\noindent where $\mathcal{E}_0$ is the amplitude of each linearly polarized centrifuge arm $E_{x,y}$, $\omega _0$ is the central frequency, and $\varphi(t)=\beta t^2$ is the quadratic phase defining the linear frequency chirp $\beta $ (notice the opposite sign of chirp for the two arms). The centrifuge field at the output of the quarter-wave plate can then be calculated as a product of $\mathbf{E}_0$ and a Jones matrix $\hat{M}_\alpha$ describing the wave plate oriented at an angle $\alpha $ \cite{HechtBook}:
\begin{align}\label{eq-Ecfg}
&\mathbf{E}_\text{cfg} = \hat{M}_\alpha \times \mathbf{E}_0 = \frac{\mathcal{E}_0 e^{-i\omega_0 t}}{\sqrt{2}} \times\\
&\left\{ \left[ \begin{array}{c} e^{-i\varphi(t)} \\ e^{+i\varphi(t)} \end{array} \right]
- i \left[ \begin{array}{c} \cos (2\alpha) e^{-i\varphi(t)} + \sin (2\alpha) e^{i\varphi(t)} \\
\sin (2\alpha) e^{-i\varphi(t)} - \cos (2\alpha) e^{i\varphi(t)} \end{array} \right] \right\}. \nonumber
\end{align}
\noindent Finally, the torque $\mathbf{T}$ on the right hand side of Eq.(\ref{eq-Euler}) exerted on a molecule by the centrifuge field is expressed as:
\begin{equation}\label{eq-torque}
\tilde{\mathbf{T}}=\langle \tilde{\mathbf{d}} \times \tilde{\mathbf{E}}_\text{cfg} \rangle=\langle \hat{\bm{\alpha}} \tilde{\mathbf{E}}_\text{cfg} \times \tilde{\mathbf{E}}_\text{cfg} \rangle,
\end{equation}
\noindent where $\mathbf{d}$ is the centrifuge-induced electric dipole moment, $\hat{\bm{\alpha}}$ is a tensor of molecular polarizability (not to be confused with the wave plate angle $\alpha $), and $\langle..\rangle$ denotes time averaging over the optical period. The \textit{tilde} sign above vectors $\mathbf{d}$, $\mathbf{T}$ and $\mathbf{E}_\text{cfg}$ indicates their conversion to the molecular frame by means of the quaternion algebra \cite{Tutunnikov2018}.

The results of our numerical simulations for the centrifuge intensity of $4.5\times 10^{12}$ W/cm$^{2}$ are shown with the dashed lines in Fig.~\ref{Fig-populations}(\textbf{b}). Here, the relative amounts of positive and negative super-rotors, as well as their total number $SR_\text{total}$, are normalized to the value of $SR_\text{total}$ at $\alpha =45$~degrees. The two dashed curves, labeled $SR_+$ and $SR_-$ and connecting at $\alpha =0$, reproduce well the experimentally observed amount of molecules rotating both in the centrifuge direction and against it (corresponding solid lines). The total calculated number of centrifuged molecules is also in good agreement with our experimental results (blue dashed vs solid curves labeled $SR_\text{total}$). One can see that in the range of the quarter-wave plate angles between 0 and 15~degrees, where the centrifuge field is substantially elliptical, the total number of rotating molecules remains at an almost constant level of $\approx$80\% with respect to those created by the conventional linear centrifuge ($\alpha =45$~deg).

The experimental results plotted with solid lines in panel (\textbf{b}) of Fig.~\ref{Fig-populations} correspond to the highest centrifuge intensity of $\approx 5\times 10^{12}$~W/cm$^{2}$, which we could achieve without entering the regime of the detrimental strong-field effects, such as significant ionization of the gas molecules. The latter resulted in filamentation and led to high fluctuations of our Raman signals, making the retrieval of the amount of super-rotors unreliable. Numerically simulating the effect of even stronger centrifuge with the peak intensity of $2\times 10^{13}$~W/cm$^{2}$, such as used in Ref.~\citenum{Ogden2019}, produced an interesting result. As shown by the dotted green curve at the top of Fig.~\ref{Fig-populations}(\textbf{b}), the efficiency of such a strong centrifuge in producing molecular super-rotors is \textit{increasing} with the increasing ellipticity of the centrifuge field. At $\alpha =0$~deg, equivalent to removing the quarter-wave plate altogether, the ensuing circular centrifuge creates about 15\% more molecular rotors than a conventional linear centrifuge of that intensity. Half of those molecules rotate clockwise, while the other half counter-clockwise.

The somewhat surprising result described above can be attributed to the highly nonlinear mechanism of the adiabatic spinning in the optical centrifuge. In the weak-centrifuge limit, when the number of super-rotors is rather negligible, the spinning efficiency exhibits a faster-than-linear growth with the centrifuge intensity. In the opposite strong-field regime, when the number of molecules spun by the centrifuge becomes comparable to the total number of molecules in the ensemble, the efficiency of spinning grows sub-linearly (and eventually saturates) with the increasing intensity \cite{MacPhail2020}. An elliptical centrifuge can be considered as a superposition of two conventional linear (sub-)centrifuges rotating in the opposite directions (CW and CCW). The degree of ellipticity is determined by the relative amplitudes of the counter-rotating linearly polarized fields. This picture explains the creation of both positive and negative super-rotors by the elliptical centrifuge, as revealed by our experimental and numerical results. From this picture it also follows that in the saturation regime, splitting the high intensity of a single strong centrifuge between the two counter-rotating weaker ones results in the more linear regime of operation of each centrifuge and correspondingly higher total number of super-rotors.
\begin{figure}[t!]
    \includegraphics[width=0.99\columnwidth]{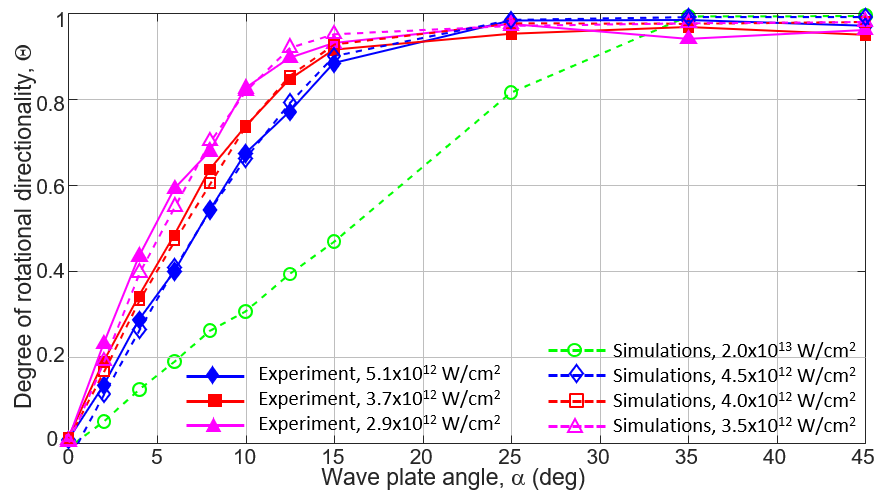}
    \caption{Experimentally measured (solid, filled markers) and numerically calculated (dashed, empty markers) degree of rotational directionality $\Theta$ as a function of the quarter-wave plate angle $\alpha $. Best fits to the experimental observations were obtained by using the centrifuge intensity as a single fitting parameter, and are plotted with the markers of the same shape as the corresponding experimental curve. In the experimental study, the intensity was limited by $\approx 5\times 10^{12}$ W/cm$^2$.}
    \label{Fig-directionality}
\end{figure}

As can be seen in Fig.~\ref{Fig-populations}, the relative number of super-rotors, rotating CW and CCW in the plane of the centrifuge, is changing gradually with the quarter-wave plate angle $\alpha $. We call the difference between the amount of positive and negative super-rotors, normalized to their total number, the degree of rotational directionality in the molecular ensemble:
\begin{equation}\label{eq-directionality}
\Theta=\frac{SR_+ - SR_-}{SR_+ + SR_-}.
\end{equation}
\noindent The dependence $\Theta(\alpha )$, extracted from our experimental observations, is shown by solid lines in Fig.~\ref{Fig-directionality}. The three curves correspond to the three highest intensities in Fig.~\ref{Fig-populations} and illustrate the main result of this study. As $\alpha $ decreases from 45 down to about 15~degrees, the level of rotational directionality is hardly changing. In this range of angles of the quarter-wave plate, the strength of the CW sub-centrifuge decreases while the total rotating field becomes more and more elliptical. As a result, it loses more and more molecules, as reflected by the decreasing number of positive super-rotors in Fig.~\ref{Fig-populations}(\textbf{b}). Yet the CCW sub-centrifuge is still too weak to capture any molecules and spin them in the opposite direction, which keeps the directionality at $\Theta\approx 1$. Only when $\alpha $ falls below 15~degrees, the re-distribution of the available photons towards the CCW sub-centrifuge makes it strong enough for creating a non-negligible amount of the negative super-rotors. From this angle and down to $\alpha =0$, the value of $SR_-$ grows at the expense of $SR_+$, and the directionality changes gradually from $\approx$1 to 0. The latter value corresponds to the circular centrifuge producing an equal amount of molecules rotating in both direction. A small disagreement between the respective values of the experimentally and numerically used intensities stems from the two main factors: (i) the nonuniform intensity distribution in the laser beam, which was not taken into account in our calculations; and (ii) the approximation of the equal weights of different Raman lines in extracting the number of super-rotors, discussed earlier in the text.

By increasing the total intensity of the centrifuge field, the controllability range of the wave plate angles can be broadened beyond the experimentally demonstrated 15~degrees. As illustrated by the green dashed line in Fig.~\ref{Fig-directionality}, a strong centrifuge at $2\times 10^{13}$~W/cm$^2$ enables a smooth variation of the degree of directionality when $\alpha $ changes between 0 and 35~degrees. In comparison to the case of lower intensities, a smaller degree of ellipticity (larger $\alpha $) is now sufficient for the CCW spinning and the correspondingly lower value of $\Theta$.

In summary, we demonstrate an experimental method of controlling the relative amount of molecules rotating clockwise and counter-clockwise with respect to a fixed laboratory axis. Such control of the rotational directionality, executed in a continuous fashion from unidirectional to bidirectional rotation, is accomplished by changing the polarization ellipticity of the laser field of an optical centrifuge by means of varying the orientation angle $\alpha $ of a single quarter-wave plate.

We study the effect both experimentally and numerically, and present a clear physical picture of the control mechanism: at any given state of the centrifuge polarization, the molecules are excited to high rotational (super-rotor) states either by climbing a single rotational ladder in the case of a conventional linearly polarized centrifuge ($|\alpha| =45$~degrees) or two rotational ladders simultaneously (when $|\alpha| < 45$~degrees). In the former case, a single rotational wave packet with $m\approx \pm J$ is created, whereas in the latter case, the centrifuge produces two distinct wave packets with $m\approx +J$ and $m\approx -J$. At $\alpha =0$~degrees, the two wave packets contain an equal amount of molecules.

Our experiments and calculations show that the degree of rotational directionality can be varied across the full range between uni- and bidirectional rotation without a significant loss in the total number of super-rotors. The latter value is determined by the intensity of the centrifuge pulse. In the strong-field limit, when the laser intensity approaches $10^{13}$~W/cm$^{2}$, the total amount of molecules centrifuged by an elliptical centrifuge remains as high as (and may even exceed) the amount of super-rotors obtained by the conventional linear centrifuge.

The demonstrated effect adds yet another ``control knob'' to the existing toolbox for harnessing molecular dynamics with laser fields. It can be useful in any studies which involve unidirectional molecular rotation, such as the centrifuge-induced molecular magnetism \cite{Milner2017a} and the enantioselective orientation of chiral molecules \cite{Milner2019}, and must be taken into account in the experiments employing optical centrifuges with nonzero polarization ellipticity.

This work was carried out under the auspices of the Canadian Center for Chirality Research on Origins and Separation (CHIROS).


\begin{thebibliography}{30}%
\makeatletter
\providecommand \@ifxundefined [1]{%
 \@ifx{#1\undefined}
}%
\providecommand \@ifnum [1]{%
 \ifnum #1\expandafter \@firstoftwo
 \else \expandafter \@secondoftwo
 \fi
}%
\providecommand \@ifx [1]{%
 \ifx #1\expandafter \@firstoftwo
 \else \expandafter \@secondoftwo
 \fi
}%
\providecommand \natexlab [1]{#1}%
\providecommand \enquote  [1]{``#1''}%
\providecommand \bibnamefont  [1]{#1}%
\providecommand \bibfnamefont [1]{#1}%
\providecommand \citenamefont [1]{#1}%
\providecommand \href@noop [0]{\@secondoftwo}%
\providecommand \href [0]{\begingroup \@sanitize@url \@href}%
\providecommand \@href[1]{\@@startlink{#1}\@@href}%
\providecommand \@@href[1]{\endgroup#1\@@endlink}%
\providecommand \@sanitize@url [0]{\catcode `\\12\catcode `\$12\catcode
  `\&12\catcode `\#12\catcode `\^12\catcode `\_12\catcode `\%12\relax}%
\providecommand \@@startlink[1]{}%
\providecommand \@@endlink[0]{}%
\providecommand \url  [0]{\begingroup\@sanitize@url \@url }%
\providecommand \@url [1]{\endgroup\@href {#1}{\urlprefix }}%
\providecommand \urlprefix  [0]{URL }%
\providecommand \Eprint [0]{\href }%
\providecommand \doibase [0]{https://doi.org/}%
\providecommand \selectlanguage [0]{\@gobble}%
\providecommand \bibinfo  [0]{\@secondoftwo}%
\providecommand \bibfield  [0]{\@secondoftwo}%
\providecommand \translation [1]{[#1]}%
\providecommand \BibitemOpen [0]{}%
\providecommand \bibitemStop [0]{}%
\providecommand \bibitemNoStop [0]{.\EOS\space}%
\providecommand \EOS [0]{\spacefactor3000\relax}%
\providecommand \BibitemShut  [1]{\csname bibitem#1\endcsname}%
\let\auto@bib@innerbib\@empty
\bibitem [{\citenamefont {Stapelfeldt}\ and\ \citenamefont
  {Seideman}(2003)}]{Stapelfeldt2003}%
  \BibitemOpen
  \bibfield  {author} {\bibinfo {author} {\bibfnamefont {H.}~\bibnamefont
  {Stapelfeldt}}\ and\ \bibinfo {author} {\bibfnamefont {T.}~\bibnamefont
  {Seideman}},\ }\href {http://link.aps.org/doi/10.1103/RevModPhys.75.543}
  {\bibfield  {journal} {\bibinfo  {journal} {Rev. Mod. Phys.}\ }\textbf
  {\bibinfo {volume} {75}},\ \bibinfo {pages} {543} (\bibinfo {year}
  {2003})}\BibitemShut {NoStop}%
\bibitem [{\citenamefont {Ohshima}\ and\ \citenamefont
  {Hasegawa}(2010)}]{Ohshima2010}%
  \BibitemOpen
  \bibfield  {author} {\bibinfo {author} {\bibfnamefont {Y.}~\bibnamefont
  {Ohshima}}\ and\ \bibinfo {author} {\bibfnamefont {H.}~\bibnamefont
  {Hasegawa}},\ }\href {https://doi.org/10.1080/0144235X.2010.511769}
  {\bibfield  {journal} {\bibinfo  {journal} {International Reviews in Physical
  Chemistry}\ }\textbf {\bibinfo {volume} {29}},\ \bibinfo {pages} {619}
  (\bibinfo {year} {2010})}\BibitemShut {NoStop}%
\bibitem [{\citenamefont {Fleischer}\ \emph {et~al.}(2012)\citenamefont
  {Fleischer}, \citenamefont {Khodorkovsky}, \citenamefont {Gershnabel},
  \citenamefont {Prior},\ and\ \citenamefont {Averbukh}}]{Fleischer2012}%
  \BibitemOpen
  \bibfield  {author} {\bibinfo {author} {\bibfnamefont {S.}~\bibnamefont
  {Fleischer}}, \bibinfo {author} {\bibfnamefont {Y.}~\bibnamefont
  {Khodorkovsky}}, \bibinfo {author} {\bibfnamefont {E.}~\bibnamefont
  {Gershnabel}}, \bibinfo {author} {\bibfnamefont {Y.}~\bibnamefont {Prior}},\
  and\ \bibinfo {author} {\bibfnamefont {I.~S.}\ \bibnamefont {Averbukh}},\
  }\href@noop {} {\bibfield  {journal} {\bibinfo  {journal} {Israel Journal of
  Chemistry}\ }\textbf {\bibinfo {volume} {52}} (\bibinfo {year}
  {2012})}\BibitemShut {NoStop}%
\bibitem [{\citenamefont {Zon}\ and\ \citenamefont
  {Katsnelson}(1975)}]{Zon1975}%
  \BibitemOpen
  \bibfield  {author} {\bibinfo {author} {\bibfnamefont {B.~A.}\ \bibnamefont
  {Zon}}\ and\ \bibinfo {author} {\bibfnamefont {B.~G.}\ \bibnamefont
  {Katsnelson}},\ }\href@noop {} {\bibfield  {journal} {\bibinfo  {journal}
  {Zh. Eksp. Teor. Fiz.}\ }\textbf {\bibinfo {volume} {69}},\ \bibinfo {pages}
  {1166} (\bibinfo {year} {1975})}\BibitemShut {NoStop}%
\bibitem [{\citenamefont {Seideman}(1995)}]{Seideman1995}%
  \BibitemOpen
  \bibfield  {author} {\bibinfo {author} {\bibfnamefont {T.}~\bibnamefont
  {Seideman}},\ }\href {http://link.aip.org/link/?JCP/103/7887/1} {\bibfield
  {journal} {\bibinfo  {journal} {The Journal of Chemical Physics}\ }\textbf
  {\bibinfo {volume} {103}},\ \bibinfo {pages} {7887} (\bibinfo {year}
  {1995})}\BibitemShut {NoStop}%
\bibitem [{\citenamefont {Friedrich}\ and\ \citenamefont
  {Herschbach}(1995)}]{Friedrich1995}%
  \BibitemOpen
  \bibfield  {author} {\bibinfo {author} {\bibfnamefont {B.}~\bibnamefont
  {Friedrich}}\ and\ \bibinfo {author} {\bibfnamefont {D.}~\bibnamefont
  {Herschbach}},\ }\href {http://link.aps.org/doi/10.1103/PhysRevLett.74.4623}
  {\bibfield  {journal} {\bibinfo  {journal} {Phys. Rev. Lett.}\ }\textbf
  {\bibinfo {volume} {74}},\ \bibinfo {pages} {4623} (\bibinfo {year}
  {1995})}\BibitemShut {NoStop}%
\bibitem [{\citenamefont {Vrakking}\ and\ \citenamefont
  {Stolte}(1997)}]{Vrakking1997}%
  \BibitemOpen
  \bibfield  {author} {\bibinfo {author} {\bibfnamefont {M.~J.~J.}\
  \bibnamefont {Vrakking}}\ and\ \bibinfo {author} {\bibfnamefont
  {S.}~\bibnamefont {Stolte}},\ }\href
  {http://www.sciencedirect.com/science/article/pii/S0009261497004363}
  {\bibfield  {journal} {\bibinfo  {journal} {Chem. Phys. Lett.}\ }\textbf
  {\bibinfo {volume} {271}},\ \bibinfo {pages} {209} (\bibinfo {year}
  {1997})}\BibitemShut {NoStop}%
\bibitem [{\citenamefont {Underwood}\ \emph {et~al.}(2005)\citenamefont
  {Underwood}, \citenamefont {Sussman},\ and\ \citenamefont
  {Stolow}}]{Underwood2005}%
  \BibitemOpen
  \bibfield  {author} {\bibinfo {author} {\bibfnamefont {J.~G.}\ \bibnamefont
  {Underwood}}, \bibinfo {author} {\bibfnamefont {B.~J.}\ \bibnamefont
  {Sussman}},\ and\ \bibinfo {author} {\bibfnamefont {A.}~\bibnamefont
  {Stolow}},\ }\href {http://link.aps.org/doi/10.1103/PhysRevLett.94.143002}
  {\bibfield  {journal} {\bibinfo  {journal} {Phys. Rev. Lett.}\ }\textbf
  {\bibinfo {volume} {94}},\ \bibinfo {pages} {143002} (\bibinfo {year}
  {2005})}\BibitemShut {NoStop}%
\bibitem [{\citenamefont {Fleischer}\ \emph {et~al.}(2009)\citenamefont
  {Fleischer}, \citenamefont {Khodorkovsky}, \citenamefont {Prior},\ and\
  \citenamefont {Averbukh}}]{Fleischer2009}%
  \BibitemOpen
  \bibfield  {author} {\bibinfo {author} {\bibfnamefont {S.}~\bibnamefont
  {Fleischer}}, \bibinfo {author} {\bibfnamefont {Y.}~\bibnamefont
  {Khodorkovsky}}, \bibinfo {author} {\bibfnamefont {Y.}~\bibnamefont
  {Prior}},\ and\ \bibinfo {author} {\bibfnamefont {I.~S.}\ \bibnamefont
  {Averbukh}},\ }\href {http://stacks.iop.org/1367-2630/11/i=10/a=105039}
  {\bibfield  {journal} {\bibinfo  {journal} {New Journal of Physics}\ }\textbf
  {\bibinfo {volume} {11}},\ \bibinfo {pages} {105039} (\bibinfo {year}
  {2009})}\BibitemShut {NoStop}%
\bibitem [{\citenamefont {Kitano}\ \emph {et~al.}(2009)\citenamefont {Kitano},
  \citenamefont {Hasegawa},\ and\ \citenamefont {Ohshima}}]{Kitano2009}%
  \BibitemOpen
  \bibfield  {author} {\bibinfo {author} {\bibfnamefont {K.}~\bibnamefont
  {Kitano}}, \bibinfo {author} {\bibfnamefont {H.}~\bibnamefont {Hasegawa}},\
  and\ \bibinfo {author} {\bibfnamefont {Y.}~\bibnamefont {Ohshima}},\ }\href
  {http://link.aps.org/doi/10.1103/PhysRevLett.103.223002} {\bibfield
  {journal} {\bibinfo  {journal} {Phys. Rev. Lett.}\ }\textbf {\bibinfo
  {volume} {103}},\ \bibinfo {pages} {223002} (\bibinfo {year}
  {2009})}\BibitemShut {NoStop}%
\bibitem [{\citenamefont {Zhdanovich}\ \emph {et~al.}(2011)\citenamefont
  {Zhdanovich}, \citenamefont {Milner}, \citenamefont {Bloomquist},
  \citenamefont {Floss}, \citenamefont {Averbukh}, \citenamefont {Hepburn},\
  and\ \citenamefont {Milner}}]{Zhdanovich2011}%
  \BibitemOpen
  \bibfield  {author} {\bibinfo {author} {\bibfnamefont {S.}~\bibnamefont
  {Zhdanovich}}, \bibinfo {author} {\bibfnamefont {A.~A.}\ \bibnamefont
  {Milner}}, \bibinfo {author} {\bibfnamefont {C.}~\bibnamefont {Bloomquist}},
  \bibinfo {author} {\bibfnamefont {J.}~\bibnamefont {Floss}}, \bibinfo
  {author} {\bibfnamefont {I.~S.}\ \bibnamefont {Averbukh}}, \bibinfo {author}
  {\bibfnamefont {J.~W.}\ \bibnamefont {Hepburn}},\ and\ \bibinfo {author}
  {\bibfnamefont {V.}~\bibnamefont {Milner}},\ }\href
  {http://link.aps.org/doi/10.1103/PhysRevLett.107.243004} {\bibfield
  {journal} {\bibinfo  {journal} {Phys. Rev. Lett.}\ }\textbf {\bibinfo
  {volume} {107}},\ \bibinfo {pages} {243004} (\bibinfo {year}
  {2011})}\BibitemShut {NoStop}%
\bibitem [{\citenamefont {Karras}\ \emph {et~al.}(2015)\citenamefont {Karras},
  \citenamefont {Ndong}, \citenamefont {Hertz}, \citenamefont {Sugny},
  \citenamefont {Billard}, \citenamefont {Lavorel},\ and\ \citenamefont
  {Faucher}}]{Karras2015}%
  \BibitemOpen
  \bibfield  {author} {\bibinfo {author} {\bibfnamefont {G.}~\bibnamefont
  {Karras}}, \bibinfo {author} {\bibfnamefont {M.}~\bibnamefont {Ndong}},
  \bibinfo {author} {\bibfnamefont {E.}~\bibnamefont {Hertz}}, \bibinfo
  {author} {\bibfnamefont {D.}~\bibnamefont {Sugny}}, \bibinfo {author}
  {\bibfnamefont {F.}~\bibnamefont {Billard}}, \bibinfo {author} {\bibfnamefont
  {B.}~\bibnamefont {Lavorel}},\ and\ \bibinfo {author} {\bibfnamefont
  {O.}~\bibnamefont {Faucher}},\ }\href
  {http://link.aps.org/doi/10.1103/PhysRevLett.114.103001} {\bibfield
  {journal} {\bibinfo  {journal} {Phys. Rev. Lett.}\ }\textbf {\bibinfo
  {volume} {114}},\ \bibinfo {pages} {103001} (\bibinfo {year}
  {2015})}\BibitemShut {NoStop}%
\bibitem [{\citenamefont {Karczmarek}\ \emph {et~al.}(1999)\citenamefont
  {Karczmarek}, \citenamefont {Wright}, \citenamefont {Corkum},\ and\
  \citenamefont {Ivanov}}]{Karczmarek1999}%
  \BibitemOpen
  \bibfield  {author} {\bibinfo {author} {\bibfnamefont {J.}~\bibnamefont
  {Karczmarek}}, \bibinfo {author} {\bibfnamefont {J.}~\bibnamefont {Wright}},
  \bibinfo {author} {\bibfnamefont {P.}~\bibnamefont {Corkum}},\ and\ \bibinfo
  {author} {\bibfnamefont {M.}~\bibnamefont {Ivanov}},\ }\href
  {http://link.aps.org/abstract/PRL/v82/p3420
  http://prl.aps.org.ezproxy.library.ubc.ca/abstract/PRL/v82/i17/p3420_1
  http://prl.aps.org.ezproxy.library.ubc.ca/pdf/PRL/v82/i17/p3420_1} {\bibfield
   {journal} {\bibinfo  {journal} {Phys. Rev. Lett.}\ }\textbf {\bibinfo
  {volume} {82}},\ \bibinfo {pages} {3420} (\bibinfo {year}
  {1999})}\BibitemShut {NoStop}%
\bibitem [{\citenamefont {Villeneuve}\ \emph {et~al.}(2000)\citenamefont
  {Villeneuve}, \citenamefont {Aseyev}, \citenamefont {Dietrich}, \citenamefont
  {Spanner}, \citenamefont {Ivanov},\ and\ \citenamefont
  {Corkum}}]{Villeneuve2000}%
  \BibitemOpen
  \bibfield  {author} {\bibinfo {author} {\bibfnamefont {D.~M.}\ \bibnamefont
  {Villeneuve}}, \bibinfo {author} {\bibfnamefont {S.~A.}\ \bibnamefont
  {Aseyev}}, \bibinfo {author} {\bibfnamefont {P.}~\bibnamefont {Dietrich}},
  \bibinfo {author} {\bibfnamefont {M.}~\bibnamefont {Spanner}}, \bibinfo
  {author} {\bibfnamefont {M.~Y.}\ \bibnamefont {Ivanov}},\ and\ \bibinfo
  {author} {\bibfnamefont {P.~B.}\ \bibnamefont {Corkum}},\ }\href
  {http://link.aps.org/abstract/PRL/v85/p542} {\bibfield  {journal} {\bibinfo
  {journal} {Phys. Rev. Lett.}\ }\textbf {\bibinfo {volume} {85}},\ \bibinfo
  {pages} {542} (\bibinfo {year} {2000})}\BibitemShut {NoStop}%
\bibitem [{\citenamefont {Yuan}\ \emph {et~al.}(2011)\citenamefont {Yuan},
  \citenamefont {Teitelbaum}, \citenamefont {Robinson},\ and\ \citenamefont
  {Mullin}}]{Yuan2011}%
  \BibitemOpen
  \bibfield  {author} {\bibinfo {author} {\bibfnamefont {L.}~\bibnamefont
  {Yuan}}, \bibinfo {author} {\bibfnamefont {S.~W.}\ \bibnamefont
  {Teitelbaum}}, \bibinfo {author} {\bibfnamefont {A.}~\bibnamefont
  {Robinson}},\ and\ \bibinfo {author} {\bibfnamefont {A.~S.}\ \bibnamefont
  {Mullin}},\ }\href {http://www.pnas.org/content/108/17/6872.abstract}
  {\bibfield  {journal} {\bibinfo  {journal} {Proceedings of the National
  Academy of Sciences}\ }\textbf {\bibinfo {volume} {108}},\ \bibinfo {pages}
  {6872} (\bibinfo {year} {2011})}\BibitemShut {NoStop}%
\bibitem [{\citenamefont {Korobenko}\ \emph {et~al.}(2014)\citenamefont
  {Korobenko}, \citenamefont {Milner},\ and\ \citenamefont
  {Milner}}]{Korobenko2014a}%
  \BibitemOpen
  \bibfield  {author} {\bibinfo {author} {\bibfnamefont {A.}~\bibnamefont
  {Korobenko}}, \bibinfo {author} {\bibfnamefont {A.~A.}\ \bibnamefont
  {Milner}},\ and\ \bibinfo {author} {\bibfnamefont {V.}~\bibnamefont
  {Milner}},\ }\href {http://link.aps.org/doi/10.1103/PhysRevLett.112.113004}
  {\bibfield  {journal} {\bibinfo  {journal} {Phys. Rev. Lett.}\ }\textbf
  {\bibinfo {volume} {112}},\ \bibinfo {pages} {113004} (\bibinfo {year}
  {2014})}\BibitemShut {NoStop}%
\bibitem [{\citenamefont {MacPhail-Bartley}\ \emph {et~al.}(2020)\citenamefont
  {MacPhail-Bartley}, \citenamefont {Wasserman}, \citenamefont {Milner},\ and\
  \citenamefont {Milner}}]{MacPhail2020}%
  \BibitemOpen
  \bibfield  {author} {\bibinfo {author} {\bibfnamefont {I.}~\bibnamefont
  {MacPhail-Bartley}}, \bibinfo {author} {\bibfnamefont {W.~W.}\ \bibnamefont
  {Wasserman}}, \bibinfo {author} {\bibfnamefont {A.~A.}\ \bibnamefont
  {Milner}},\ and\ \bibinfo {author} {\bibfnamefont {V.}~\bibnamefont
  {Milner}},\ }\href {https://doi.org/10.1063/1.5140358} {\bibfield  {journal}
  {\bibinfo  {journal} {Rev. Sci. Instrum.}\ }\textbf {\bibinfo {volume}
  {91}},\ \bibinfo {pages} {045122} (\bibinfo {year} {2020})}\BibitemShut
  {NoStop}%
\bibitem [{\citenamefont {Armon}\ and\ \citenamefont
  {Friedland}(2016)}]{Armon2016}%
  \BibitemOpen
  \bibfield  {author} {\bibinfo {author} {\bibfnamefont {T.}~\bibnamefont
  {Armon}}\ and\ \bibinfo {author} {\bibfnamefont {L.}~\bibnamefont
  {Friedland}},\ }\href {https://doi.org/10.1103/PhysRevA.93.043406} {\bibfield
   {journal} {\bibinfo  {journal} {Phys. Rev. A}\ }\textbf {\bibinfo {volume}
  {93}},\ \bibinfo {pages} {043406} (\bibinfo {year} {2016})}\BibitemShut
  {NoStop}%
\bibitem [{\citenamefont {Spanner}\ and\ \citenamefont
  {Ivanov}(2001)}]{Spanner2001}%
  \BibitemOpen
  \bibfield  {author} {\bibinfo {author} {\bibfnamefont {M.}~\bibnamefont
  {Spanner}}\ and\ \bibinfo {author} {\bibfnamefont {M.~Y.}\ \bibnamefont
  {Ivanov}},\ }\href {http://dx.doi.org/10.1063/1.1342225} {\bibfield
  {journal} {\bibinfo  {journal} {The Journal of Chemical Physics}\ }\textbf
  {\bibinfo {volume} {114}},\ \bibinfo {pages} {3456} (\bibinfo {year}
  {2001})}\BibitemShut {NoStop}%
\bibitem [{\citenamefont {Spanner}\ \emph {et~al.}(2001)\citenamefont
  {Spanner}, \citenamefont {Davitt},\ and\ \citenamefont
  {Ivanov}}]{Spanner2001a}%
  \BibitemOpen
  \bibfield  {author} {\bibinfo {author} {\bibfnamefont {M.}~\bibnamefont
  {Spanner}}, \bibinfo {author} {\bibfnamefont {K.~M.}\ \bibnamefont
  {Davitt}},\ and\ \bibinfo {author} {\bibfnamefont {M.~Y.}\ \bibnamefont
  {Ivanov}},\ }\href {http://dx.doi.org/10.1063/1.1407271} {\bibfield
  {journal} {\bibinfo  {journal} {The Journal of Chemical Physics}\ }\textbf
  {\bibinfo {volume} {115}},\ \bibinfo {pages} {8403} (\bibinfo {year}
  {2001})}\BibitemShut {NoStop}%
\bibitem [{\citenamefont {Armon}\ and\ \citenamefont
  {Friedland}(2017)}]{Armon2017}%
  \BibitemOpen
  \bibfield  {author} {\bibinfo {author} {\bibfnamefont {T.}~\bibnamefont
  {Armon}}\ and\ \bibinfo {author} {\bibfnamefont {L.}~\bibnamefont
  {Friedland}},\ }\href {https://doi.org/10.1103/PhysRevA.96.033411} {\bibfield
   {journal} {\bibinfo  {journal} {Phys. Rev. A}\ }\textbf {\bibinfo {volume}
  {96}},\ \bibinfo {pages} {1} (\bibinfo {year} {2017})}\BibitemShut {NoStop}%
\bibitem [{\citenamefont {Milner}\ \emph {et~al.}(2016)\citenamefont {Milner},
  \citenamefont {Korobenko},\ and\ \citenamefont {Milner}}]{Milner2016b}%
  \BibitemOpen
  \bibfield  {author} {\bibinfo {author} {\bibfnamefont {A.~A.}\ \bibnamefont
  {Milner}}, \bibinfo {author} {\bibfnamefont {A.}~\bibnamefont {Korobenko}},\
  and\ \bibinfo {author} {\bibfnamefont {V.}~\bibnamefont {Milner}},\ }\href
  {http://link.aps.org/doi/10.1103/PhysRevA.93.053408} {\bibfield  {journal}
  {\bibinfo  {journal} {Phys. Rev. A}\ }\textbf {\bibinfo {volume} {93}},\
  \bibinfo {pages} {053408} (\bibinfo {year} {2016})}\BibitemShut {NoStop}%
\bibitem [{\citenamefont {Ogden}\ \emph {et~al.}(2019)\citenamefont {Ogden},
  \citenamefont {Michael}, \citenamefont {Murray}, \citenamefont {Liu},
  \citenamefont {Toro},\ and\ \citenamefont {Mullin}}]{Ogden2019}%
  \BibitemOpen
  \bibfield  {author} {\bibinfo {author} {\bibfnamefont {H.~M.}\ \bibnamefont
  {Ogden}}, \bibinfo {author} {\bibfnamefont {T.~J.}\ \bibnamefont {Michael}},
  \bibinfo {author} {\bibfnamefont {M.~J.}\ \bibnamefont {Murray}}, \bibinfo
  {author} {\bibfnamefont {Q.}~\bibnamefont {Liu}}, \bibinfo {author}
  {\bibfnamefont {C.}~\bibnamefont {Toro}},\ and\ \bibinfo {author}
  {\bibfnamefont {A.~S.}\ \bibnamefont {Mullin}},\ }\href
  {https://doi.org/10.1039/C8CP06917D} {\bibfield  {journal} {\bibinfo
  {journal} {Physical Chemistry Chemical Physics}\ }\textbf {\bibinfo {volume}
  {21}},\ \bibinfo {pages} {14103} (\bibinfo {year} {2019})}\BibitemShut
  {NoStop}%
\bibitem [{\citenamefont {Korech}\ \emph {et~al.}(2013)\citenamefont {Korech},
  \citenamefont {Steinitz}, \citenamefont {Gordon}, \citenamefont {Averbukh},\
  and\ \citenamefont {Prior}}]{Korech2013}%
  \BibitemOpen
  \bibfield  {author} {\bibinfo {author} {\bibfnamefont {O.}~\bibnamefont
  {Korech}}, \bibinfo {author} {\bibfnamefont {U.}~\bibnamefont {Steinitz}},
  \bibinfo {author} {\bibfnamefont {R.~J.}\ \bibnamefont {Gordon}}, \bibinfo
  {author} {\bibfnamefont {I.~S.}\ \bibnamefont {Averbukh}},\ and\ \bibinfo
  {author} {\bibfnamefont {Y.}~\bibnamefont {Prior}},\ }\href
  {https://doi.org/10.1038/nphoton.2013.189
  http://www.nature.com/nphoton/journal/v7/n9/abs/nphoton.2013.189.html#supplementary-information}
  {\bibfield  {journal} {\bibinfo  {journal} {Nat. Photon.}\ }\textbf {\bibinfo
  {volume} {7}},\ \bibinfo {pages} {711} (\bibinfo {year} {2013})}\BibitemShut
  {NoStop}%
\bibitem [{\citenamefont {Bitter}\ and\ \citenamefont
  {Milner}(2016)}]{Bitter2016c}%
  \BibitemOpen
  \bibfield  {author} {\bibinfo {author} {\bibfnamefont {M.}~\bibnamefont
  {Bitter}}\ and\ \bibinfo {author} {\bibfnamefont {V.}~\bibnamefont
  {Milner}},\ }\href {http://link.aps.org/doi/10.1103/PhysRevLett.117.144104}
  {\bibfield  {journal} {\bibinfo  {journal} {Phys. Rev. Lett.}\ }\textbf
  {\bibinfo {volume} {117}},\ \bibinfo {pages} {144104} (\bibinfo {year}
  {2016})}\BibitemShut {NoStop}%
\bibitem [{\citenamefont {Seideman}\ and\ \citenamefont
  {Hamilton}(2005)}]{Seideman2005}%
  \BibitemOpen
  \bibfield  {author} {\bibinfo {author} {\bibfnamefont {T.}~\bibnamefont
  {Seideman}}\ and\ \bibinfo {author} {\bibfnamefont {E.}~\bibnamefont
  {Hamilton}},\ }\bibinfo {title} {Nonadiabatic alignment by intense pulses.
  concepts, theory, and directions},\ in\ \href
  {https://doi.org/http://dx.doi.org/10.1016/S1049-250X(05)52006-8} {\emph
  {\bibinfo {booktitle} {Advances In Atomic, Molecular, and Optical
  Physics}}},\ Vol.\ \bibinfo {volume} {Volume 52},\ \bibinfo {editor} {edited
  by\ \bibinfo {editor} {\bibfnamefont {P.~R.}\ \bibnamefont {Berman}}\ and\
  \bibinfo {editor} {\bibfnamefont {C.~C.}\ \bibnamefont {Lin}}}\ (\bibinfo
  {publisher} {Academic Press},\ \bibinfo {year} {2005})\ pp.\ \bibinfo {pages}
  {289--329}\BibitemShut {NoStop}%
\bibitem [{\citenamefont {Tutunnikov}\ \emph {et~al.}(2018)\citenamefont
  {Tutunnikov}, \citenamefont {Gershnabel}, \citenamefont {Gold},\ and\
  \citenamefont {Averbukh}}]{Tutunnikov2018}%
  \BibitemOpen
  \bibfield  {author} {\bibinfo {author} {\bibfnamefont {I.}~\bibnamefont
  {Tutunnikov}}, \bibinfo {author} {\bibfnamefont {E.}~\bibnamefont
  {Gershnabel}}, \bibinfo {author} {\bibfnamefont {S.}~\bibnamefont {Gold}},\
  and\ \bibinfo {author} {\bibfnamefont {I.~S.}\ \bibnamefont {Averbukh}},\
  }\href {https://doi.org/10.1021/acs.jpclett.7b03416} {\bibfield  {journal}
  {\bibinfo  {journal} {The Journal of Physical Chemistry Letters}\ }\textbf
  {\bibinfo {volume} {9}},\ \bibinfo {pages} {1105} (\bibinfo {year}
  {2018})}\BibitemShut {NoStop}%
\bibitem [{\citenamefont {Hecht}(2002)}]{HechtBook}%
  \BibitemOpen
  \bibfield  {author} {\bibinfo {author} {\bibfnamefont {E.}~\bibnamefont
  {Hecht}},\ }\href@noop {} {\emph {\bibinfo {title} {Optics}}},\ \bibinfo
  {edition} {5th}\ ed.\ (\bibinfo  {publisher} {Addison Wesley},\ \bibinfo
  {address} {San Francisco},\ \bibinfo {year} {2002})\BibitemShut {NoStop}%
\bibitem [{\citenamefont {Milner}\ \emph {et~al.}(2017)\citenamefont {Milner},
  \citenamefont {Korobenko},\ and\ \citenamefont {Milner}}]{Milner2017a}%
  \BibitemOpen
  \bibfield  {author} {\bibinfo {author} {\bibfnamefont {A.~A.}\ \bibnamefont
  {Milner}}, \bibinfo {author} {\bibfnamefont {A.}~\bibnamefont {Korobenko}},\
  and\ \bibinfo {author} {\bibfnamefont {V.}~\bibnamefont {Milner}},\ }\href
  {https://link.aps.org/doi/10.1103/PhysRevLett.118.243201} {\bibfield
  {journal} {\bibinfo  {journal} {Phys. Rev. Lett.}\ }\textbf {\bibinfo
  {volume} {118}},\ \bibinfo {pages} {243201} (\bibinfo {year}
  {2017})}\BibitemShut {NoStop}%
\bibitem [{\citenamefont {Milner}\ \emph {et~al.}(2019)\citenamefont {Milner},
  \citenamefont {Fordyce}, \citenamefont {MacPhail-Bartley}, \citenamefont
  {Wasserman}, \citenamefont {Milner}, \citenamefont {Tutunnikov},\ and\
  \citenamefont {Averbukh}}]{Milner2019}%
  \BibitemOpen
  \bibfield  {author} {\bibinfo {author} {\bibfnamefont {A.~A.}\ \bibnamefont
  {Milner}}, \bibinfo {author} {\bibfnamefont {J.~A.~M.}\ \bibnamefont
  {Fordyce}}, \bibinfo {author} {\bibfnamefont {I.}~\bibnamefont
  {MacPhail-Bartley}}, \bibinfo {author} {\bibfnamefont {W.}~\bibnamefont
  {Wasserman}}, \bibinfo {author} {\bibfnamefont {V.}~\bibnamefont {Milner}},
  \bibinfo {author} {\bibfnamefont {I.}~\bibnamefont {Tutunnikov}},\ and\
  \bibinfo {author} {\bibfnamefont {I.~S.}\ \bibnamefont {Averbukh}},\ }\href
  {https://doi.org/10.1103/PhysRevLett.122.223201} {\bibfield  {journal}
  {\bibinfo  {journal} {Phys. Rev. Lett.}\ }\textbf {\bibinfo {volume} {122}},\
  \bibinfo {pages} {223201} (\bibinfo {year} {2019})}\BibitemShut {NoStop}%
\end{thebibliography}

%

\end{document}